# A REVIEW: STUDY OF HANDOVER PERFORMANCE IN MOBILE IP

Geetanjali Chellani, Anshuman Kalla

Department of Electronics and Communication Engineering
Jaipur National University, Jaipur, Rajasthan, India

## *ABSTRACT*

*The Mobile Internet Protocol (Mobile IP) is an extension to the Internet Protocol proposed by the Internet Engineering Task Force (IETF) that addresses the mobility issues. In order to support un-interrupted services and seamless mobility of nodes across the networks (and/or sub-networks) with permanent IP addresses, handover is performed in mobile IP enabled networks. Handover in mobile IP is source cause of performance degradation as it results in increased latency and packet loss during handover. Other issues like scalability issues, ordered packet delivery issues, control plane management issues etc are also adversely affected by it. The paper provides a constructive survey by classifying, discussing and comparing different handover techniques that have been proposed so far, for enhancing the performance during handovers. Finally some general solutions that have been used to solve handover related problems are briefly discussed.*

## 1. INTRODUCTION

Foundation of today's Internet architecture, based on TCP/IP, was laid during the days of telephony when enabling communication between static end users was of prime importance and mobility of users was least envisioned. But the advent of wireless technology gave rise to the possibility of mobility and seamless connectivity. Among the several other solutions that have been proposed so far, Mobile IP is the only widely deployed add-on solution for handling mobility[1]. In the TCP/IP based Internet architecture a user node is assigned an IP address which is in-fact a locator of user-node in network. As the node moves from one place to another, it results in change of network and/or subnet which consequently results in change of IP address. Since all the connections take IP address as a seed thus change in IP address means that all the connections must be re-established which inevitably leads to interruption in on-going applications and services. This issue of varying IP addresses when node is mobile is resolved by Mobile IP. There are still some issues that need to be reconsidered looking at the enormous growth of mobile users every-day-and-now. As discussed by J. Chandrasekarn [2] these issues are (i) Handover Latency, (ii) Triangulation, (iii) Reliability and (iv) Security. In this paper we will discuss all these issues.

The organization of the paper is as follow. Section II introduces current solutions for supporting mobility in IPv4 & IPv6 and major differences between them. In section III, mobility management and related components is presented. The network mobility for mobile networks is introduced in section IV. In section V different handover techniques that improve handover performance are discussed and a comparative study is done. Section VI introduces some general techniques that are used to improve handover performance followed by the conclusion in section VII.

## 2. MOBILE IP

Mobile IPv4 (MIPv4) is popular mobility internet protocol used in different IPv4 networks and Mobile IPv6 (MIPv6) has emerged to deal with mobility for advanced version of IP i.e. IPv6.

### 2.1. MOBILE IPV4

Mobile IPv4 introduced four functional entities: (i) Home Agent (HA), (ii) Foreign Agent (FA), (iii) Mobile Node (MN), (iv) Correspondent Node (CN). Each MN is resident in its home network where it receives a permanent Home Address (HoA).When an MN moves out of its home network and visits a foreign network, it obtains a temporary address which is known as Care-of-Address (CoA) by the FA in that foreign network. When the MN moves from one foreign network to another foreign network, it registers its new CoA to the HA that is located in the home network. The HA keeps track of the HoA and CoA for all MN. A packet from CN destined to MN is sent to HoA of MN. The HA intercepts all the IP packets destined to the MN and tunnels them to the CoA of the MN [6].

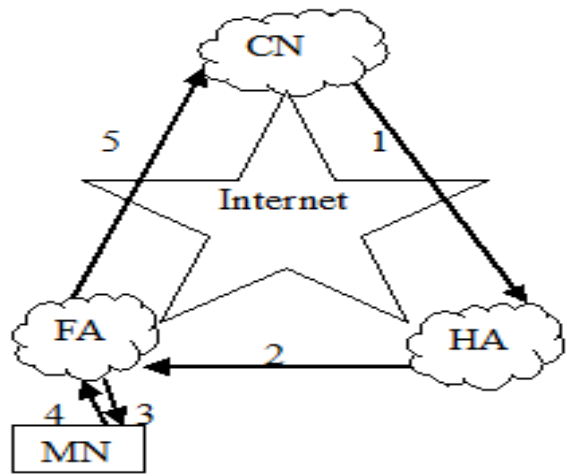

Fig.1 Mobile IPv4

#### 2.1.1. Basic Mobile IPv4 Protocols Functioning

**2.1.1.1. Agent Discovery** - In order to discover prevailing agent i.e. home agent or foreign agent, a mobile node invokes this mechanism. Two different types of messages used are:

**2.1.1.1.1. Agent Advertisement** - Home/foreign agent advertises its presence periodically by broadcasting agent advertisement message with-in its network.

**2.1.1.1.2. Agent Solicitation** - MN can also issue a request message with-in the current network in order to seek an agent advertisement message.

**2.1.1.2. Registration** - Mobile node visiting a foreign network informs about its current location by initiating a registration procedure.

**2.1.1.2.1. Registration Request** - Visiting mobile node after fetching temporary CoA from foreign network needs to convey this CoA to the home agent so it generates and sends a registration request message.

**2.1.1.2.2. Registration Reply** - Upon reception of registration request, home agent verifies the authenticity of mobile node. In case of authentic request, a mapping of CoA is established with corresponding HoA of mobile node, by adding an entry in routing table. Finally an acknowledgment is sent to MN in form of registration reply message.

**2.1.1.3. Tunneling** - Tunneling is used to forward IP datagram from a home address to a care of address.

**2.1.2. Issues in Mobile IPv4**

**2.1.2.1. Triangular Routing** - Mobile IPv4 suffers from a long handover delay due to "triangular routing". As shown in figure 1, packets going from MN to CN follow direct route through internet (i.e.4&5) but packets going from CN to the MN have to travel through HA when the mobile node is away from home (i.e.1,2&3). This additional routing is called triangular routing.

**2.1.2.2. Signalling Overhead** - Large signalling overhead is due to large number of registration updates. Every time a mobile node moves beyond the limit of link layer connectivity, a registration update is required for the node with its home agent [7].

**2.1.3. Solution of Mobile IPv4**
Route optimization [8] was proposed to solve triangular routing problem. Messages from the CN are routed directly to the MN's CoA without passing through the HoA. The CN maintains a binding cache that maps the HoA of the mobile node with their CoA. Binding cache needs four additional messages which are as follows[9]:

**2.1.3.1. Binding Request** - In order to know the current location of MN, CN sends a binding request to HA at home network.

**2.1.3.2. Binding Update** - HA replies to CN with a message that revels the current location of an MN.

**2.1.3.3. Binding Acknowledgement** - CN acknowledges HA, the reception of binding update.

**2.1.3.4. Binding Warning** - This message is used to suggest a MN's home agent that CN appears to have either no binding cache entry or an out-of-date binding cache entry for some MN.

**2.2. Mobile IPv6** - Mobile IPv6 (MIPv6) is the next generation internet protocol and offers a number of improvements over MIPv4. MIPv6 supports mobility in both homogeneous (from one LAN to another LAN) and heterogeneous media (node movement from LAN to 3G network). In MIPv6, MN should assign three IPv6 addresses (i) Permanent home address, (ii) Current link local address, (iii) Care-of-Address (CoA), which associated with the mobile node only when visiting a particular foreign network [10]. MN's CoA is co-located CoA in MIPv6 which allow MN to encapsulate and decapsulate packets and connect to HA directly on any foreign link without notifying FA. The FA function is not there in MIPv6. While the MN moves from one network (or subnet) to another, CoA is automatically allocated to it in the foreign network due to the address auto-configuration feature which are (i) Statefull Address Auto-configuration - MN sends a CoA Request message to the local router and it allocates a new IPv6 address (ii) Stateless Address Auto-configuration - MN combines IPv6-prefix which it received with its MAC address

139

to create new IPv6 address using neighbour discovery. The HA keeps a binding between MN's HoA and its CoA. The central data structure collected by each IPv6 node is used as a binding cache. In MIPv6 route optimization is in-built function so MN periodically sends binding update messages not only to the HA but also to CN. So, CN adds this binding to the binding cache and thereafter CN directly sends packets directly to MN's CoA indicated in the binding. In MIPv6, DAD (Duplicate Address Detection) procedure is invoked to determine the uniqueness of the new MN's CoA in which a MN sends a neighbour solicitation message with a set timer to ask that this address is being used or not. If no node replies with-in the set time then MN can assume that this address is unique in that network and it could use this address.

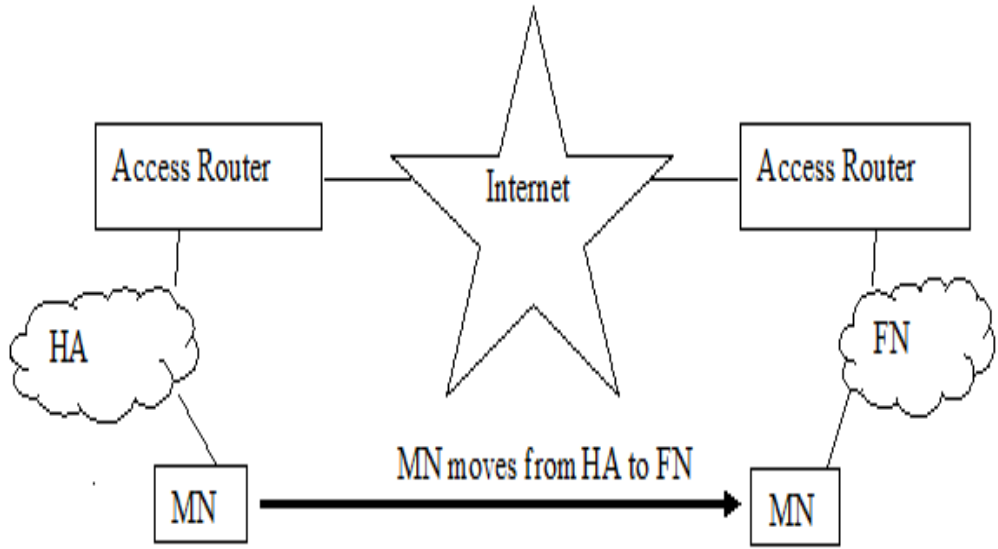

Fig. 2 Mobile IPv6

## 2.3. Distinction Between MIPv4 and MIPv6

1. Route Optimization process is a fundamental operation in MIPv6. In MIPv4, this feature is an extension which may not be supported by all nodes.

2. Address Auto-configuration is also basic part of the MIPv6 which leads to removal of FA which is used in MIPv4.

3. Packets are tunnelled using a routing header in MIPv6 where as MIPv4 uses IP encapsulation for all packets. Using routing header reduces overhead which requires less additional header bytes to be added to a packet at the time of sending packets.

4. Security is the prime concern in MIPv6 which utilizes IP Security (IPsec), where as MIPv4 utilizes mobility security association and relies on its own security mechanism for all these activities [11].

## 3. MOBILITY MANAGEMENT

Different components for mobility management are as follows:

**3.1. Handover Management** - Mobility support handover management reduces the service interruption during the handover. In Mobile IP handover latency represent the time between the last received packets from the old network until the first received packet from the new network [12]. In case of high handover latency, large number of packet could be lost. Packet losses could cause critical disruption for real time services. Thus packets should be routed with low latency as possible by IP routing and thereby alleviating packet loss during handover [13].

**3.2. Location Management** - Location management is done by the network to find out the current mobile node's location and keep tracking its movement by using movement detection algorithm[14]. Movement detection algorithms have a role of optimizing Mobile-IP handover by reducing the registration delay. In Mobile-IP there are two types of movement detection algorithms:

**3.2.1. Advertisement Based Algorithm(ABS)**[15] - This depends on periodic broadcasts from mobility agents. ABS has two distinct algorithms are:

**3.2.1.1. Lazy Cell Switching (LCS)** expects that movement of MN is rare and thus it avoids handover until it is absolutely necessary. Consequently LCS is always slow to adapt the mobility.

**3.2.1.2. Eager Cell Switching (ECS)** assumes frequent location changes and perform immediate handover upon discovering a mobility agent thereby making movement detection time negligible. Accordingly it is fast to adapt mobility.

**3.2.2. Hint Based Algorithm(HBA)**[14] - It requires information from the link layer termed as hints in order to perform movement detection.HBA has two distinct algorithms are:

**3.2.2.1. Hinted Cell Switching (HCS)** is proposed to extend the amount of information communicated from the link layer to MIP and to include information about the environment as identity of the local mobility agent. So it reduces movement detection time and Mobile-IP handover delay.

**3.2.2.2. Fast Hinted Cell Switching (FHCS)** allows link layer to send triggers to network layer whenever handover occurs. So it is able to reduce handover latency by denying the need for movement detection and identity of local mobility agent.

**3.3. Multihoming** - Multihoming is a special case of a mobility management in which the mobile device can use many access networks for example GPRS and Wi-Fi to access the internet and switch the network while moving[16]. Multihomed Mobile-IP provide MN to register multiple CoA at the HA to achieve more reliable connectivity.

**3.4. Security** - Security needs are getting active attention as wireless environment is potentially more vulnerable to attacks including passive eavesdropping, active reply attacks, insider attack and Denial of Service (DoS) attacks [2] based on the Mobile-IP registration protocol. So key management is strongly desired in order to preclude aforementioned attacks. In Mobile IPv4 mobility security association is considered while Mobile IPv6 uses IPsec.

## 4. NETWORK MOBILITY

Network Mobility (NEMO) is proposed to support mobility in mobile networks[17]. Two aspects of mobile networks are host mobility and network mobility. Host mobility has a scope of only single node which is connected and network mobility is concerned with entire network. NEMO introduces an important device termed as Mobile Router (MR) which acts as a gateway for the mobile networks to configure a connection to the mobile nodes. The mobile nodes are (i) Local Fixed Nodes (LFN) which cannot move and have the same home agent as the MR has, (ii) Local Mobile Nodes (LMN) which can move and belong to the mobile network as its home network, (iii) Visiting Mobile Nodes (VMN) which do not belong to the mobile network and attached to the mobile network as a temporary basis. IETF standard for NEMO is NEMO Basic Support Protocol (BSP), has advantages like reduce signalling and increased manageability, but also have disadvantages like inefficient route and increased handover latency. To solve the limitations of the NEMO BSP a set of NEMO Route optimization schemes are introduced[18]. Route Optimization (RO) is a solution for providing improved end-to-end path between CN and MN, reduce signalling overhead and packet loss. In [19], number of RO schemes have been introduced to overcome aforementioned disadvantages.

**4.1. Delegation** - In this RO scheme, prefix of the foreign network is delegated inside the mobile network. Mobile Network Nodes (MNNs) obtain their CoAs from received prefixes. Then the obtain CoA which send BUs (Binding Updates) to HAs and CNs. Therefore CNs have BU of MNN's CoAs, so packets are sent directly to the foreign network without considering HAs. Delegation based approach provides optimal route with low header overhead[20].

**4.2. Hierarchical** - In this scheme a packet reaches the foreign network either from MNN's HA or carried through HA of MNNs and Top Level Mobile Router (TLMR)[21]. Packets sent by CN to MNN, using MNN's HoA, reaches MNN's HA that tunnel packets to TLMR's CoA or HoA. Thus packets which are tunneled using CoA will directly go to corresponding foreign network whereas the packets, which are tunneled to HoA will go to the TLMR's HA and further TLMR sends them to MNN using MRs that maintain a routing table which contains MNN's prefix. In this scheme one tunnel always exists between the TLMR and VMN's HA, so it reduces signaling and is easily deployable.

**4.3. Source Routing** - RO has been achieved through CN by inserting CoAs of MRs in the packet header itself so that each packet knows the underlying network structure made-up of MRs. Packet are sent from CN to TLMR without going through HAs using CoA of MRs which lies in packet header, thanks to source routing. In this scheme memory requirement is low but header overhead is increased[22].

**4.4. BGP Assisted** - This scheme of RO is originated in Border Gateway Protocol (BGP), in this scheme BGP routers are always updated by using forwarding entries for the prefix of the mobile network in the routing table when the mobile network moves. This information about the mobile network moves, is flagged to few routers that swap the information containing routing entries to forward packets to the mobile network with each other using routing protocol through internet[23]. Signaling is reduced but scalability is increased in maintaining routing entries.

## 5. ENHANCED HANDOVER SCHEMES IN MIPV6

**Handover Delay** - Handover delay is considered as time taken to redirect the on-going communication from previous to current point-of-attachment[3]. Moreover handover delay is composed of two types of delays. (i) Registration Delay - This delay is considered as time taken during the HA registration process, (ii) Resolution Delay - This delay is considered as time taken

International Journal of Computer Networks & Communications (IJCNC) Vol.5, No.6, November 2013

[29], when MN configure a new location CoA, if it is in the foreign network. To overcome these delays which provided interruption in communication many handover approaches has been proposed by several authors which are described in next sections. Also a comparison between these techniques based on many factors like handover latency, packet loss, signaling overhead etc is presented at the end.

**5.1. Hierarchical Mobile IP (HMIP)[24]** - To address the problem of HA registration delay, many hierarchical networks have been proposed in which internet is separated into different administrative domains. Movement of MN with a single administrative domain is called micro mobility while movement across different administrative domains is called macro mobility. In hierarchical Mobile-IP Mobility Anchor Point (MAP) is used as a router that maintains the binding process for the mobile nodes currently visiting its domain. The MAP is considered as a HA of the MN. MAP intercepts the packets targeted to the MNs addresses inside the domain and then tunnels them to the correspondent CoA of the MNs in their foreign network. When MN moves inside the domain it register their CoA to MAP only, there is no need to inform the HA so this is called Local Care of Address (LCoA) for inside domain movement but when MN moves to a new MAP domain it obtains Regional Care of Address (RCoA) for outside domain movement [25]. After obtaining address the MN sends a binding update to the MAP which will bind the MN's RCoA to its LCoA. MAP then sends binding acknowledgement to MN for informing successful registration. One more binding update is sent to MN's HA when MN changes the entire MAP domain. So such network reduces signaling overhead as well as handover delay by reducing home agent registration when MN moves inside the domain[26][37].

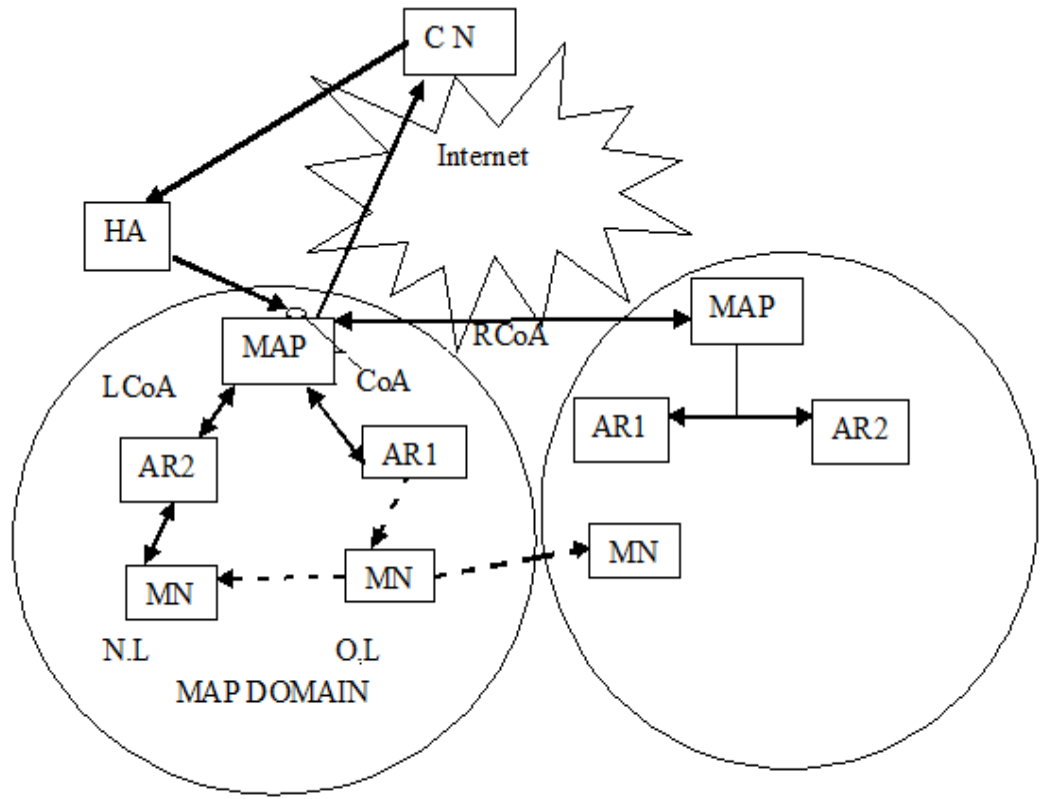

Fig.3. Hierarchical MIP

Various HMIP based mobility protocols are; (i) Paging HMIPv6 (PHMIP) [27], proposes paging services in MAP domain which provides information to MAP domain about the MN when it moves in an in-active mode (no active communication session) and determine exact location of MN using paging criteria thus it reduces power consumption, (ii) Robust HMIPV6 (RHMIP) [28], MN registers with two different MAPs known as Primary MAP (P-MAP) and Secondary MAP (S-MAP) simultaneously. When MN or CN detects a failure of P-MAP, it changes its attachment from P-MAP to S-MAP. Hence it is more robustness and resilient by improving failure recovery time. On the other hand it results in increased signaling overhead, (iii) Mobility Based Binding Update HMIPV6 (MBBUHMIP) [29], provides lifetime value of binding cache and introduceslocation update of MN by reducing signaling cost, (iv) Multilevel Hierarchy HMIPV6 (MHHMIP) [30], uses tree structure hierarchy of MAP thus providing scalable service but suffers from extra packet processing overhead, (v) FF-HMIP (FF-HMIP) [31], based on HMIP prevents global handover signaling by appointing a MAP and uses a fast MIP that reduces handover latency by link layer trigger. Hence, it achieves improved handover performance and signaling overhead at the cost of additional tunneling header, (vi) HMIP over Multiprotocol Label Switching (HMIP-MPLS) [32], provides mobility and multimedia service by merging radio access network with HMIP without any alteration in HMIP protocol, so signaling overhead is increased due to two merged protocols. Summary of comparison is given in Table-1.

Table 1. Comparison between significant techniques based on HMIP [33]

| HMIPv6 Protocol | Vantages | Drawback | Signaling Overhead |
|---|---|---|---|
| PHMIP (2003) | Signaling overhead and power consumption is reduced. | Increase handover latency due to inter domain movement | Low |
| RHMIP (2003) | Obtain robustness and fault tolerance. | Signaling overhead due to multiple registration | High |
| MBBUHMIP (2003) | Reduce signaling overhead adjusted MN's lifetime using MN's mobility pattern. | Increase binding update and signaling cost | Low |
| MHHMIP (2004) | Supports multi level hierarchal structure. | Packet processing and signaling overhead | High |
| FF-HMIP (2004) | Improves signaling overhead and handover performance. | Introduce tunneling overhead | Low |
| HMIP-MPLS (2007) | Supports multilevel protocol switching over HMIPv6. | Additional signaling overhead | High |

**5.2. Fast Handover Mobile IP (FHMIP)** - To address the problem of FA address resolution delay, FHMIP has been proposed in which MN will pre-configure a new CoA when it moves from old Access Router (oAR) to new Access Router (nAR). It has three different types are:

**5.2.1. MN initiated handover** – When fast handover is about to occur, it is MN that gets first notification from link layer (L2) information. Accordingly MN sends a Router Solicitation for Proxy (RtSolPr) message to oAR as well as to new access node. Along with RtSolPr message MN send sent link layer address to new access node. In response, oAR sends the Proxy Router Advertisement (PrRtAdv) message to MN, which provides information about the new access node that includes link-layer address and prefixes. On receiving PrRtAdv message MN decides a prospective CoA based on prefix of selected nAR. Further MN sends Fast-Binding Update (FBU) to the oAR and in response oAR sends Handover Initiation (HI) message to nAR for imminent handover [34][35]. After that nAR returns a Handover Acknowledgement (HAck) message to oAR in order to establish a binding between old CoA (oCoA) to new CoA (nCoA). In

144

response of HAck, oAR sends duple Fast Binding Acknowledgement (F-BAck) to MN and nAR for forwarding the MN's traffic towards nCoA. The nAR buffers the packets until MN establishes link connectivity with the nAR. The MN sends a Fast Neighbor Advertisement (F-NA) to inform the nAR of its presence and finally nAR sends the buffered packets to MN.

**5.2.2. Network Initiated Handover** - In such type of handovers, networks are made capable of initiating handovers. However process of message exchanging is slightly different. PrRtAdv message is sent by an oAR in an unsolicited way to the MN which contain the information (configuring CoA) about the new networks in absence of initial RtSolPr message[36].

**5.2.3. Reactive Handover** – Unlike MN initiated and network initiated handovers, the oAR does not receive FBU from MN before connectivity ends. Therefore HI, Hack and F-BAck messages are not present. MN sends FBU to nAR by encapsulating with the Fast Neighbor Advertisement (FNA) message. Further, nAR send this FBU to oAR. oAR then allows to create a binding between oCoA and nCoA. Further the oAR forwards the MN's traffic to the nAR and nAR in term send the traffic to MN. FHMIP uses wireless link layer (L2) trigger based information for smoothing of handover procedure and minimizing the FA resolution delay [37].

Some of the important research done over FHMIP are discussed below: (i) Fast MIPv6 (FMIPv6) [38], provides seamless handover by making use of layer-2 trigger to obtain new link CoA while still being connected to the previous link in order to reduce packet loss, (ii) Simultaneously Binding Fast Handover (SBFHMIPv6) [39], provides simultaneous binding function at the MN. MN's traffic is multi casted to current location as well as to the locations where MN could roam in near future, (iii)Seamless Multicasting Fast Handover (SMFHMIPv6) [40],provides integrated unicast and multicast handover with combination of fast handover that creates seamless multicast handover, [41], (iv) Pre-Binding Fast Handover (PBFHMIPv6) presents a modified version of FMIPv6 using extra binding updates such as pre-binding update and pre-binding acknowledgement between nAR and oAR. Thus there is no need to established reverse tunneling between nAR and oAR, (v) Early Binding Fast Handover (EBFHMIPv6) [42], provides EBFH in which an MN completes its binding update with current access router before link-going-down trigger (i.e. MN is close to handover), (vi) Simplified Fast Handover (SFHMIPv6) [43], significantly increases the probability that the protocol can successfully perform the fast handover procedure in predictive mode which MN cannot complete due to lack of time in FMIPv6 version. SFHMIPv6 also reduces anticipation time. A tabular summary is given in Table- 2.

Table 2. Comparison between significant techniques based on FHMIP

| FHMIPv6 Protocol | Vantages | Drawback | Handover Delay |
|---|---|---|---|
| FMIPv6(2005) | MN perform fast handover in predictive mode, So no packet loss | Additional signaling overhead due to additional signaling message are required for handover | High |
| SBFHMIPv6(2006) | Provides simultaneous binding to reduce packet loss | Protocol enables to decouple L2 and L3 handover, so signaling overhead | High |
| SMFMIPv6(2006) | Packet processing overhead reduce due to air interface | Additional signaling message | Still High |
| PBFHMIPv6(2006) | Remove tunneling | Extra binding update create signaling overhead | High |
| EBFHMIPv6(2006) | Provides fast handover for fast moving nodes | Consumes large amount of network performance and creates overhead | Comparative Low |
| SFHMIPv6(2008) | Supporting high speed MN movement in predictive mode | Reduce signaling cost and packet delivery cost | Low |

**5.3. Seamless Handover Mobile IP (SH-MIP)** -Seamless handover is an improved version handover which is based on hierarchical network and fast handover. The main aim of this handover scheme is to reduce packet loss by using Synchronized Packet Simulcast (SPS) (packets are broadcast on both oAR and nAR) and hybrid handover mechanism (tracking of MN's current location and its signal strength). In seamless Mobile-IP a new entity introduced is Design Engine (DE) which have mainly four functions, (i) To control handover process, (ii) Take decision for handovers at the network domain, (iii) Keeps location tracking of all the mobile nodes by identifying the movement modes (linearly, stochastically, stationary), (iv) Offers load balancing when MN connects with lower load access routers [44]. The seamless handover occurs when MN wants to go to a new network. When MN receives beacon advertisement message from adjacent nAR, then it sends RtSolPr message to oAR for initiating the handover. oAR then sends HI message to adjacent nAR which contain nCoA and oCoA. In response, HAck is send by the nAR to oAR for establishing a binding between oCoA to nCoA. Further oAR sends Carrying Load State (CLS) message to DE periodically which indicates the number of MN's related to the AR and their IP addresses. MN also sends Current Tracking State (CTS) message to DE periodically when it receives beacon advertisement message from nAR which indicates the signal strength of nAR. After determining CLS and CTS messages and tracking the mobile node movement, DE sends Handover Decision (HD) to all ARs, following which oAR sends Handover Notification (HN) message (which is extracted from HD) to MN that indicates the MN to which nAR it must handover. In response, MN sends F-BU to oAR in order to bind its link address with nCoA, after that oAR send Simulcast (Scast) message to MAP which initiates simulcasting of packets (i.e. duplication) and sending the packets to oAR and nAR's cache buffer at the same time. oAR and MAP sends F-BAck to both current and new networks for ensuring reception of its message. MN sends F-NA message to nAR when it connects to the new link and nAR forwards packets to MN. At same time oAR also forwards the packets to nAR. On completion of packet sending from oAR to MN through MAP, nAR sends Simulcast off (Soff) message to the MAP and MAP forwards this message to DE which indicates that MN does not execute another seamless handover process until current handover process is not completed.

Significant work has been done an SH-MIP, some of them are summarized here: (i) Adaptive SH over video streaming (ASHMIPv6-VS) [45], presents an adaptive mobile video streaming scheme for dynamically establishing network conditions. MN always buffers frames for disruption in connectivity during handover so it is easy for streaming media server to adapt the video being streamed to MN during handover to support seamless mobility, (ii) SH for Proxy MobileIPV6 (SH-PMIPV6) [46], it is a network based approach to control mobility management on behalf of the MN so that MN is not required in order to provide any information about the target network, (iii) Optimized SHMIPv6 (OSHMIPv6) [47], uses dynamic distributed algorithm which belongs to the b-matching problem to select regional MAP that achieves peer-to-peer communication mode in handover process, (iv) SHMIPv6 based on cellular network (SH-CN) [48], allows MN to utilize their oCoA on the new link. It provides not only expedited forwarding of packets to MN but also accelerated forwarding packets to their correspondents, (v) SH for IP Multimedia Subsystem over MobileIPv6 (IMS-SHMIPv6) [49], presents context transfer mechanisms based on predictive and reactive schemes. It also provides QoS provisioning for improvement of the service quality of IP Multimedia Subsystem(IMS), (vi) Secure Password Authentication Mechanism for SHPMIPv6 (SPAM-SHPMIPv6) [50], introduces a modified version of SH-PMIPv6 that provides high security, resists various attacks (forgery attack, reply attack, stolen verified attack) and performs authentication procedure by using bi-casting scheme based on piggy-backing technique to reduce packet loss. Table- 3 provides comparative summary of all techniques under SHMI. Finally a comparison between all the broad categories is presented in Table- 4.

International Journal of Computer Networks & Communications (IJCNC) Vol.5, No.6, November 2013

Table 3. Comparison between significant techniques based on SHMIP

| SHMIPv6 Protocol | Vantages | Drawbacks | Handover Delay |
|---|---|---|---|
| ASHMIPv6-VS(2006) | Support cross layer approach to adapt the changes in the network condition | Extra frames are buffered | Handover delay is minimized |
| SH-PMIPv6(2008) | Avoids on-the-fly packet loss while ensuring the packet sequence | Suffers from packet buffering in order to perform packet ordering | By using neighbor discovery message handover latency is reduced |
| OSHMIPv6(2011) | Reduce packet loss and improve handover performance | Signaling cost is increased | Reduce handoff latency |
| SH-CN(2011) | Provides Pre-configure bi directional secure tunnels to accelerate mobility management | Introduce tunneling key overhead | Handover delay is reduced in both inter domain and intra domain movements |
| IMS-SHMIPv6(2012) | Introduce IP multimedia subsystem for real time application | Introduce signaling message overhead for re-register and re-invite of MN for re-establishment of the session | Reduce handover latency |
| SPAM-SHMIPv6(2013) | Avoiding packet loss problem and reduces signaling overhead | Memory requirement is increased due to buffering | Handover latency is minimized |

Table 4. Comparison between significant Handover Techniques

| Handover schemes | Handover Latency | Packet Loss | Signaling overhead | Route Optimization | Deployment | Packet Buffering |
|---|---|---|---|---|---|---|
| MIPv4 | Long | High | High | Optional process | FA is deployed in MIPv4 | No |
| MIPv6 | Lengthy handover delay | Moderate | High | In-built process | No extra functional component is used | No |
| HMIPv6 | Moderate | High | Low due to signaling overhead | In-built process | Gateway foreign agent and Regional foreign agent is used | Yes |
| FH-MIPv6 | Low using L2 trigger restricted under movement speed of the MN | Moderate | Low | In-built process | No extra functional component is used | Yes |
| Handover schemes | Handover Latency | Packet Loss | Signaling overhead | Route Optimization | Deployment | Packet Buffering |
| SH-MIPv6 | Low | Low | Low | In-built process | Design Engine is used | Yes |

147

## 6. GENERAL SOLUTIONS

**6.1. Buffering** - Buffering is a general solution deployed for avoiding packet loss during handover. In this scheme when handover occurs, all the packets which are destined to previous FA for the MN are forwarded to new FA by using buffering which happens by notifying the CoA of new FA [51].

**6.2. HAWAII** - Handoff Aware Wireless Access Internet Infrastructure (HAWAII) is a solution for improving handover latency. It is very similar to the hierarchal network but in this scheme packets are routed in intra-domain and route optimization strategies are also implemented in order to reduce handover delay [52].

**6.3. Exclusive Handover Message (EHM)** - EHM is another solution for improving packet loss. This scheme gives end-to-end approach and improves the bust effect of host mobility on TCP performance in wireless network. It calculates timeout at the Base Station (BS) when handover occurs. This information about the handover is easily acquired by receiving router advertisement occurs. This information about the handover is easily acquired by receiving router advertisement message from new base station so BS sends EHM to fixed node to avoid retransmission of packets at fixed node[53].

**6.4. Mobile IP Fast Authentication Protocol (MIFA)[54]** - MIFA is used to solve handover latency. This scheme is based on local authentication with the new FA and independent of re-authorization with the HA. MIFA uses security associations like MN-HA, MN-FA which adds extra security between the connections which enables the FA to authenticate the MN. Unlike hierarchical MIP it does not require hierarchical of FA's.

**6.5. Enhanced Mobile IP (E-MIP)** - E-MIP is a solution for improving handover latency and packet loss. It improves handovers through link layer information which allows an MN to predict the loss of connectivity before connection is lost. A forceful handover is made to new network even before any mobility is detected at network layer [55]. So it reduces handover latency by eliminating the time required for handover detection at the network layer when mobility occurs. So fast and seamless handover is achieved if MN's moving speed is not high or not low .

## 7. SUMMARY

The luxury of seamless connectivity and interruption free access to the internet anytime and anywhere to users requires network to ensure that mobile node remains attached with globally known permanent IP address even on a move and packets are delivered correctly without loss during transit. An overview and comparative study of Hierarchical Mobile IP, Fast handover, Seamless handover is presented. The global aim of all techniques is to remove packet loss, end to end delay, handover latency and signaling load resulting in smooth handover.

International Journal of Computer Networks & Communications (IJCNC) Vol.5, No.6, November 2013

**AUTHORS**

**Geetanjali Chellani** She completed her B.Tech Degree in Electronics and Communication Engineering from Rajasthan Technical University, Kota in year 2010. She is pursing M.Tech in Communication and Signal Processing from Jaipur National University, Jaipur, Rajasthan, India. Her areas of interest includes Wireless Networking, Micro-controller, Digital Signal Processing, Digital Electronics, Circuit Analysis.

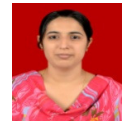

**Anshuman Kalla** is at present working as an Assistant Professor at department of Electronics and Communication Engineering, Jaipur National University. He did Bachelor's of Engineering (B.E.) from Engineering College Bikaner (Rajasthan University) in 2004. He has pursued two funded Masters; First from ISEP, Paris, France in 2008 and second from University of Nice Sophia Antipolis, France in 2011. In addition, he has completed two research based internship one at Alcatel Lucent Technologies and another at Orange Labs, France. He has worked on Genetic Algorithms and its implementation in networks, Peer-to-peer video streaming, Delay Tolerant Network and Content Centric Networking.

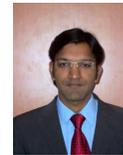